\documentclass{PoS}

\usepackage[round]{natbib}

\newcommand{\sub}[1]{_{\rm #1}}
\def\farcs{\hbox{$.\!\!^{\prime\prime}$}}

\title{Imaging pulsar echoes at low frequencies}

\ShortTitle{Imaging pulsar echoes at low frequencies}

\author{\speaker{Olaf Wucknitz}\\ Max-Planck-Institut f\"ur
  Radioastronomie, Auf dem H\"ugel 69, 53121 Bonn, Germany,\\ E-mail:
  \email{wucknitz@mpifr-bonn.mpg.de}}

\abstract{Interstellar scattering is known to broaden distant objects
  spatially and temporally. The latter aspect is difficult to analyse,
  unless the signals carry their own time stamps. Pulsars are so kind
  to do us this favour. Typically the signature is a broadened image
  with little or no substructure and a similarly smooth exponential
  scattering tail in the temporal profile. The case of the pulsar
  B1508+55 is special: The profile shows additional components that
  are moving relative to the main pulse with time. We use
  low-frequency VLBI with LOFAR to test the hypothesis that these
  components are actually such scattering-induced echoes, by trying to
  detect the expected angular offset.  Using international stations
  (plus the Kilpisj{\"a}rvi Atmospheric Imaging Receiver Array
  `KAIRA') and the phased-up core of the LOFAR array, we can do
  interferometry at high resolution in time and space.

  This contribution presents a selection of results from an ongoing
  large-scale monitoring campaign. We can not only detect the offset,
  but even image a full string of echoes, and relate the positions
  with delays. What we find is apparently consistent with scattering
  by highly aligned components in a single screen at a distance of
  120\,pc. Further investigations will improve our understanding of
  the scattering process as basis of using the scattering-induced
  subimages as arms of a giant interstellar interferometer with
  insanely high resolution.  }

\FullConference{14th European VLBI Network Symposium \& Users Meeting
  (EVN 2018)\\ 8-11 October 2018\\ Granada, Spain}

\begin{document}

\section{The ghost of B1508+55}

The pulsar B1508+55 has $S\sim 800\,$mJy near 150\,MHz, which makes it
an easy target for LOFAR \citep{haarlem13}.  Monitoring revealed
interesting features (Verbiest, priv.~comm.): Donner found significant
DM variations that were followed up by Os\l{}owski, using monitoring
with GLOW stations.\footnote{The `German Long Wavelength Consortium'
  (GLOW) coordinates local use of the German LOFAR stations.}
Variability of the trailing edge of the profile were found in these
data and independently in a monitoring campaign by Serylak. Macquart
and Bhat hypothesised that the `ghost' component(s) are likely caused
by interstellar scattering and are not intrinsic to the pulsar itself.

\begin{figure}[hbt]
  \hspace*{2.7em}\includegraphics[height=0.35\textwidth]{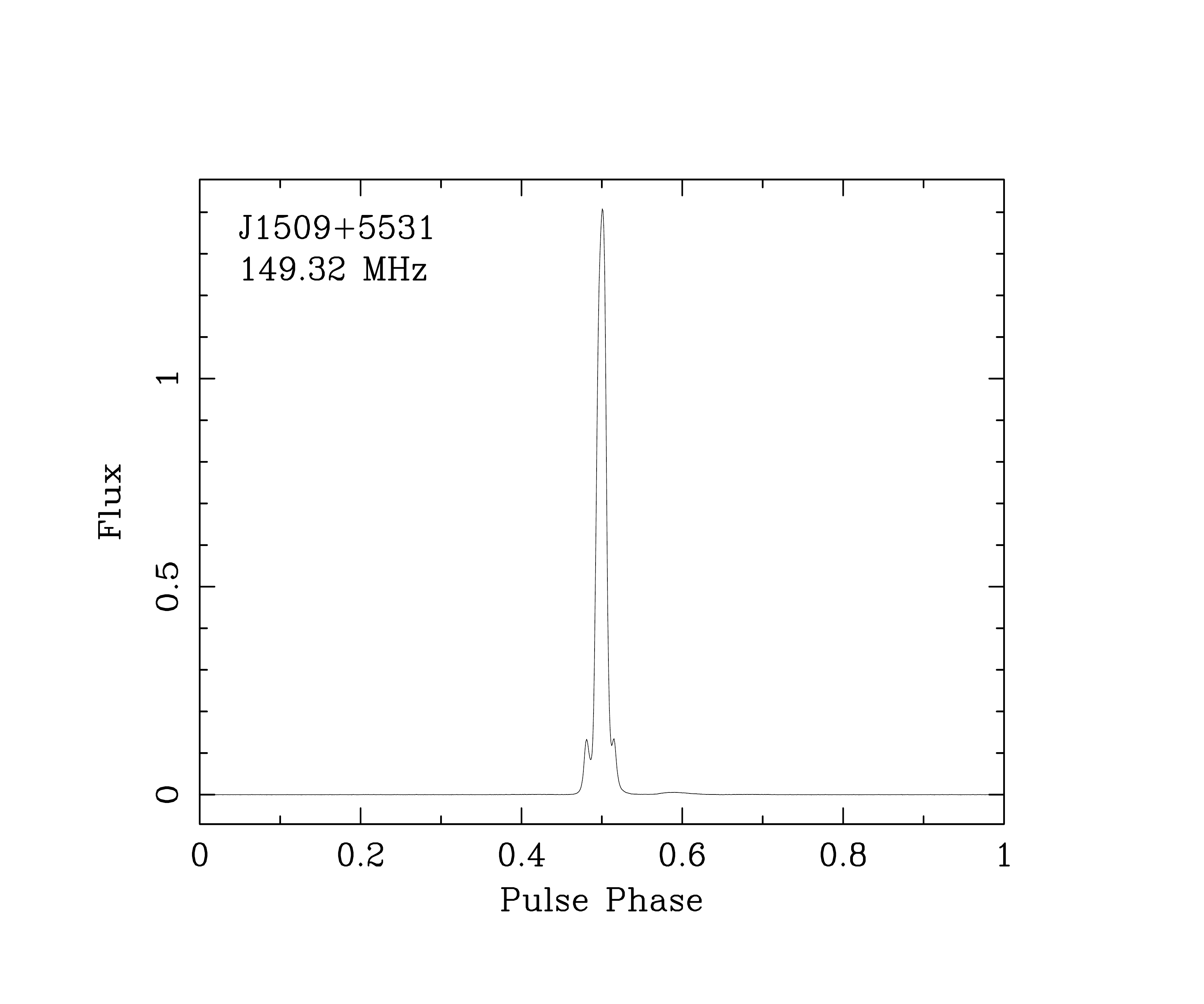}
  \includegraphics[height=0.35\textwidth]{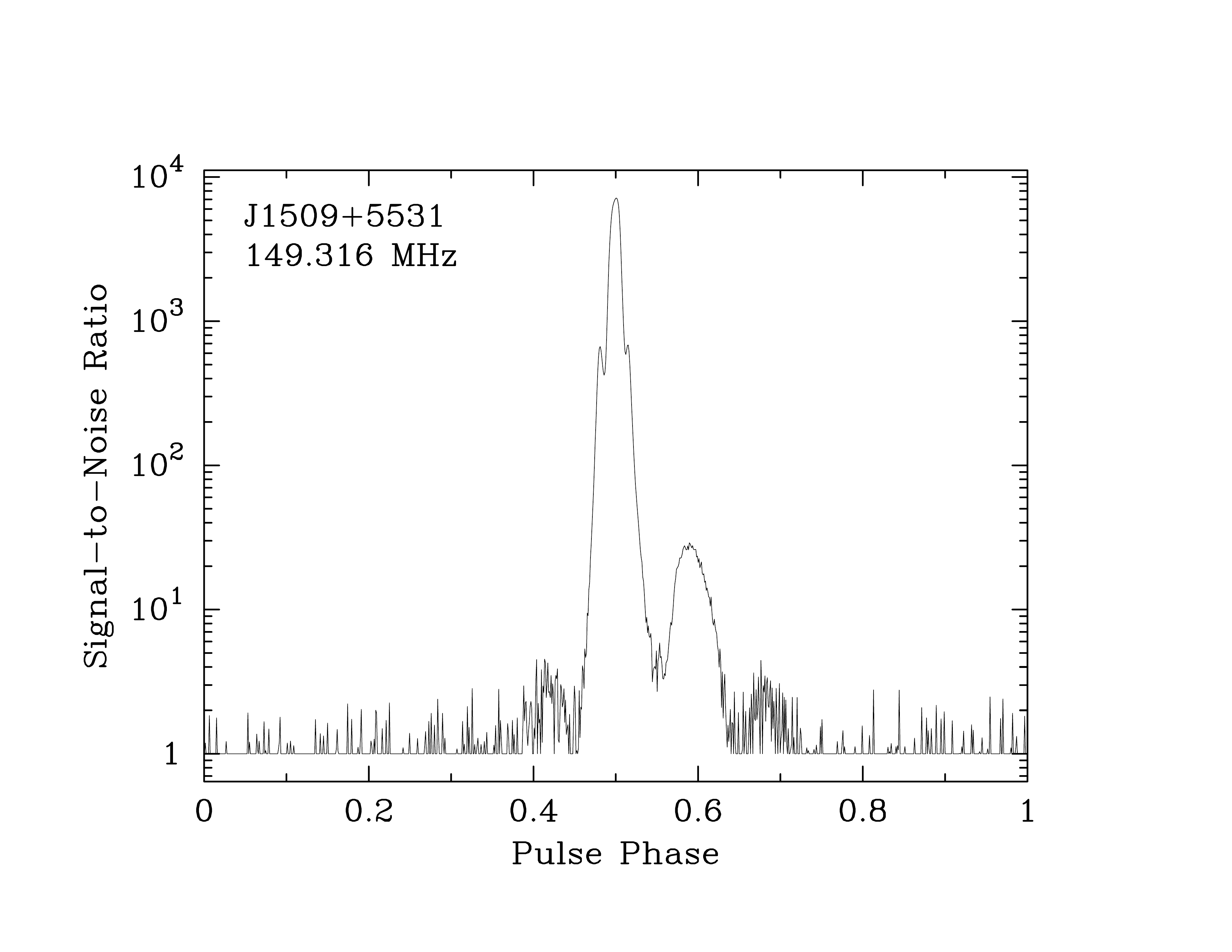}
  \caption{Profile of B1508+55 observed with LOFAR.  The ghost near
    phase 0.6 (about 70\,msec offset from the peak) becomes obvious on
    a logarithmic scale (right). Profiles kindly provided by Stefan
    Os\l{}owski.}
  \label{fig:profile}
\end{figure}

\begin{figure}[hbt]
\centering\includegraphics[width=0.6\textwidth]{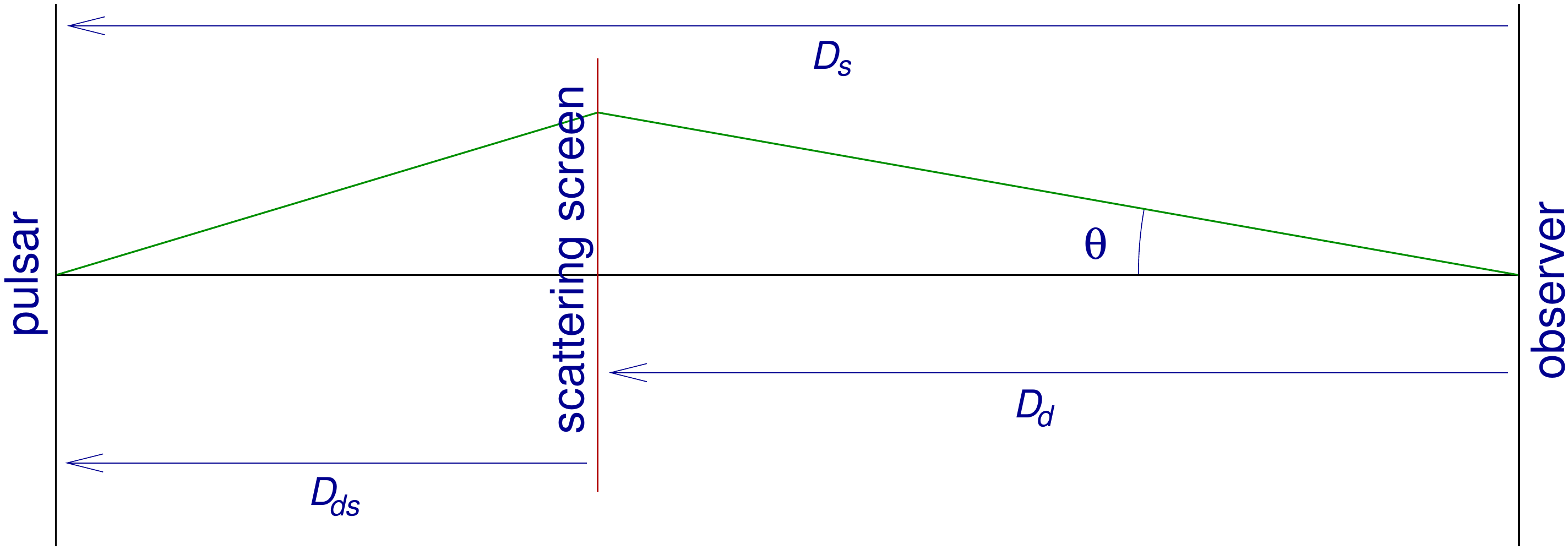}
\caption{Scattering geometry with one subimage deflected by an
  apparent angle $\theta$.}
\label{fig:geometry}
\end{figure}

Extensive LOFAR monitoring confirmed the existence of the ghost
components (see Fig.~\ref{fig:profile}) and found that they are slowly
approaching the main pulse with time.  Scattered components are
delayed, because their geometric path is slightly longer (see
Fig.~\ref{fig:geometry}).
\begin{displaymath}
c\tau = \frac12 \theta^2 D
\hspace{8em} D =\frac{D\sub{s}\,D\sub{d}}{D\sub{ds}}
\end{displaymath}
$D\sub{s}$ and $D\sub{d}$ are the distances to the source and the
deflector, respectively, and $D\sub{ds}$ is the distance between
deflector and source. For non-cosmological distances they are
additive, $D\sub{ds}=D\sub{s}-D\sub{d}$.

For a first guess we assumed that the scattering screen is halfway
between the pulsar and us, which corresponds to $D=D\sub{s}=2.1\,$kpc
\citep{chatterjee09}. For a delay of 70\,msec (see
Fig.~\ref{fig:profile}) the expected angle is $\theta=0\farcs16$. This
cannot be \emph{resolved} with LOFAR at 150\,MHz, but components
separated by pulsar gating can be \emph{localised} even better,
provided the signal is strong enough.

\section{Time-resolved VLBI with LOFAR}

We have to correlate offline after recording baseband data because of
the insufficient temporal resolution of the online correlator. This
procedure is unusual for LOFAR, but not for VLBI.  We started the
observations in 2016 with the six German stations of the LOFAR array,
which can be controlled and recorded centrally in Bonn and
J\"ulich. Later we included all other international stations in
France, Sweden, England, Poland (3), Ireland, the phased-up core and
KAIRA in Northern Finland \citep{mckay15}. The core is recorded
centrally, but all other stations use their own recording
equipment. Data are transferred electronically to Bonn.\footnote{For
  KAIRA this involves a physical transport of disks from the site to
  Troms\o{}. During winter and spring this can delay the data delivery
  significantly.} The data rate is 3.1\,Gbps per station, one hour
with all stations amounts to ca.~20\,TB of baseband data.

Data are correlated with an own software correlator. The main pulse is
used as in-beam calibrator, so that positions and fluxes are relative
to the unscattered pulsar itself. After correcting for possible
amplitude and phase differences between the linear dipoles per
station, we convert the signal to a circular polarisation basis using
an approximated model of the instrument response.

The phases are calibrated with a station-based fringe-fitting
algorithm that includes dispersive (ionospheric) and non-dispersive
(clocks) delays, rates and differential Faraday rotation. Afterwards
the amplitude bandpass is determined and corrected, and a second
iteration of fringe-fitting is performed. The calibration solutions
are then applied to the full set of time-resolved visibilities.

\begin{figure}[b]
  \hspace*{1.5em}\includegraphics[height=0.5244\textwidth]{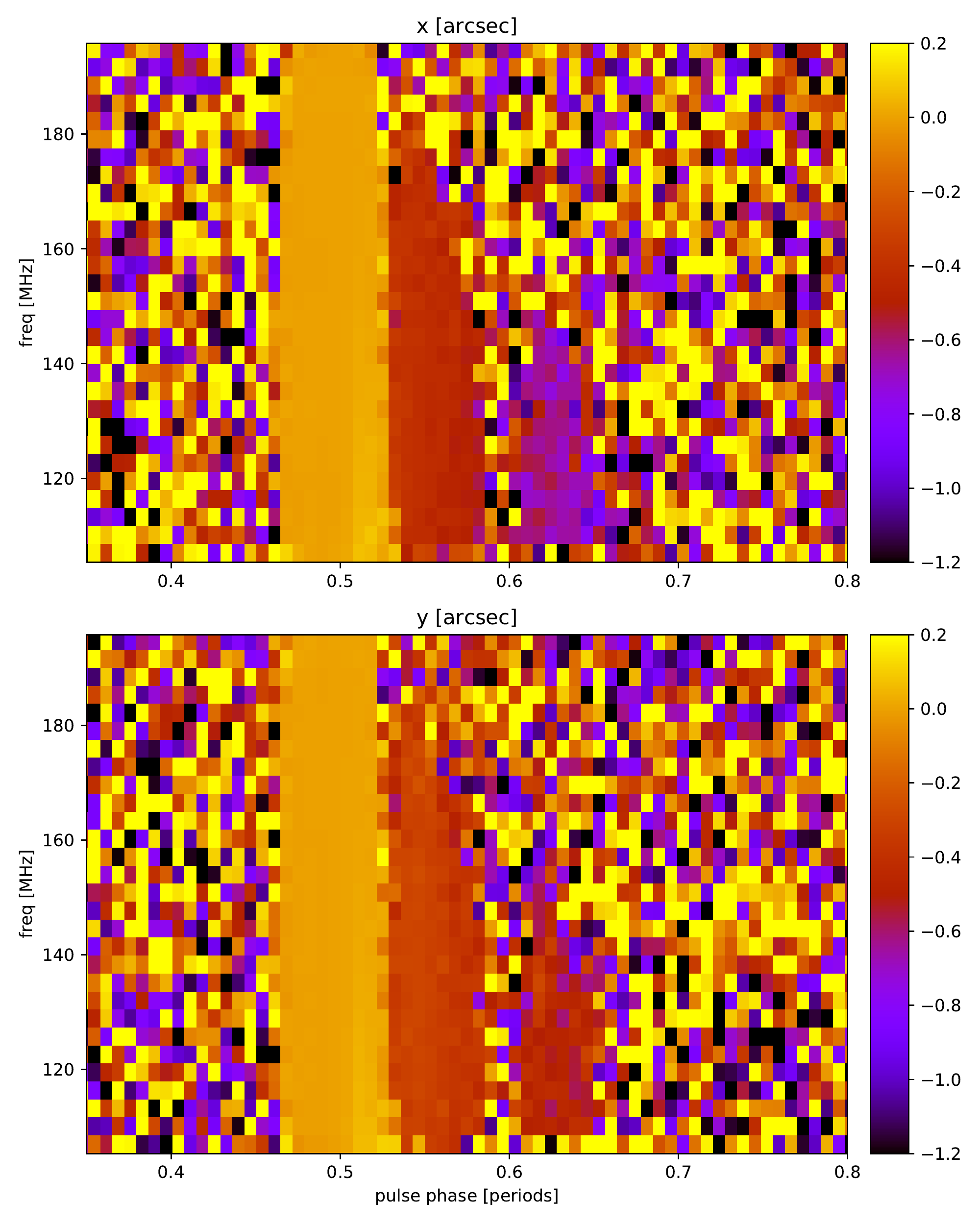}
  \hfill
  \includegraphics[height=0.51775\textwidth]{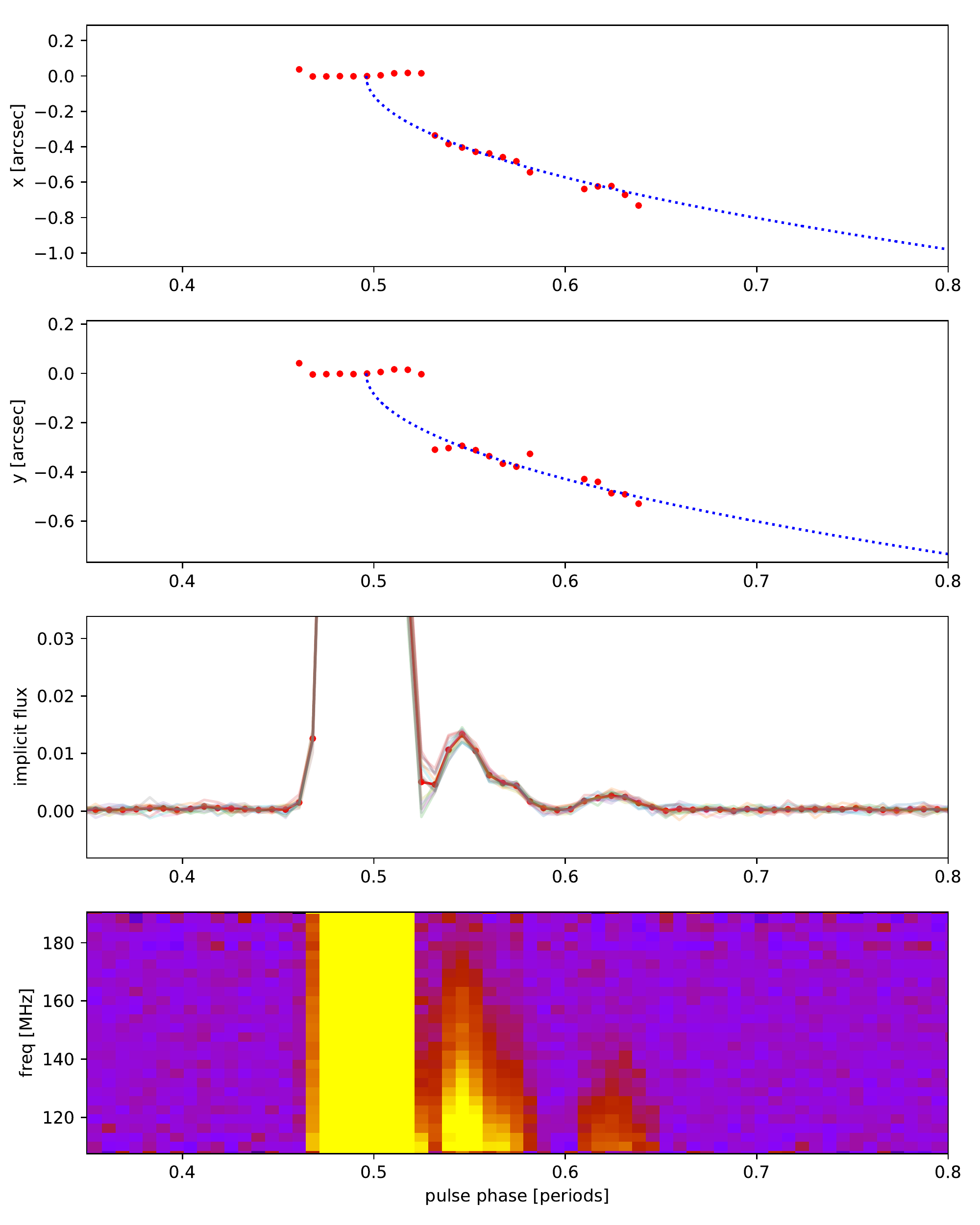}\hspace*{1.5em}
  \caption{Left: Component position as function of pulse phase and
    frequency. Right: Component position (upper two panels); flux
    (third panel, combined and single baselines) as function of pulse
    phase; folded dynamic spectrum (bottom).  The offset is consistent
    with the expected parabola. We also notice the steep spectrum of
    the echoes, which is consistent with scattering.}
  \label{fig:glow folded}
\end{figure}

Data weights are derived from the autocorrelations, which
automatically downweights visibilities affected by RFI, so that no
flagging is needed. This is remarkable given the extremely strong
interference at certain frequencies and times. A disadvantage of this
approach is that very bright pulses are downweighted slightly,
whenever they reach a significant fraction of the system noise.

\section{Gated analysis}

After the phase and amplitude calibration we combine all baselines to
measure the position and flux as function of pulse phase (proxy for
delay) and frequency. Figure~\ref{fig:glow folded} shows clearly that
the echoes are offset in position from the main pulse, which confirms
the echo hypothesis. The effect is stronger than expected, which means
the scattering medium must be closer to us.  The positions are
constant over the band so that we can combine all frequencies. The
position relative to the main pulse follows the expected parabola from
the relation $\tau\propto\theta^2$ quite well. From the slope of that
parabola we get a first estimate of the distance to the scattering
screen of approximately 124\,pc.

\begin{figure}[h]
\centering \includegraphics[width=0.29\textwidth,bb=80 18 560
  410,clip=true]{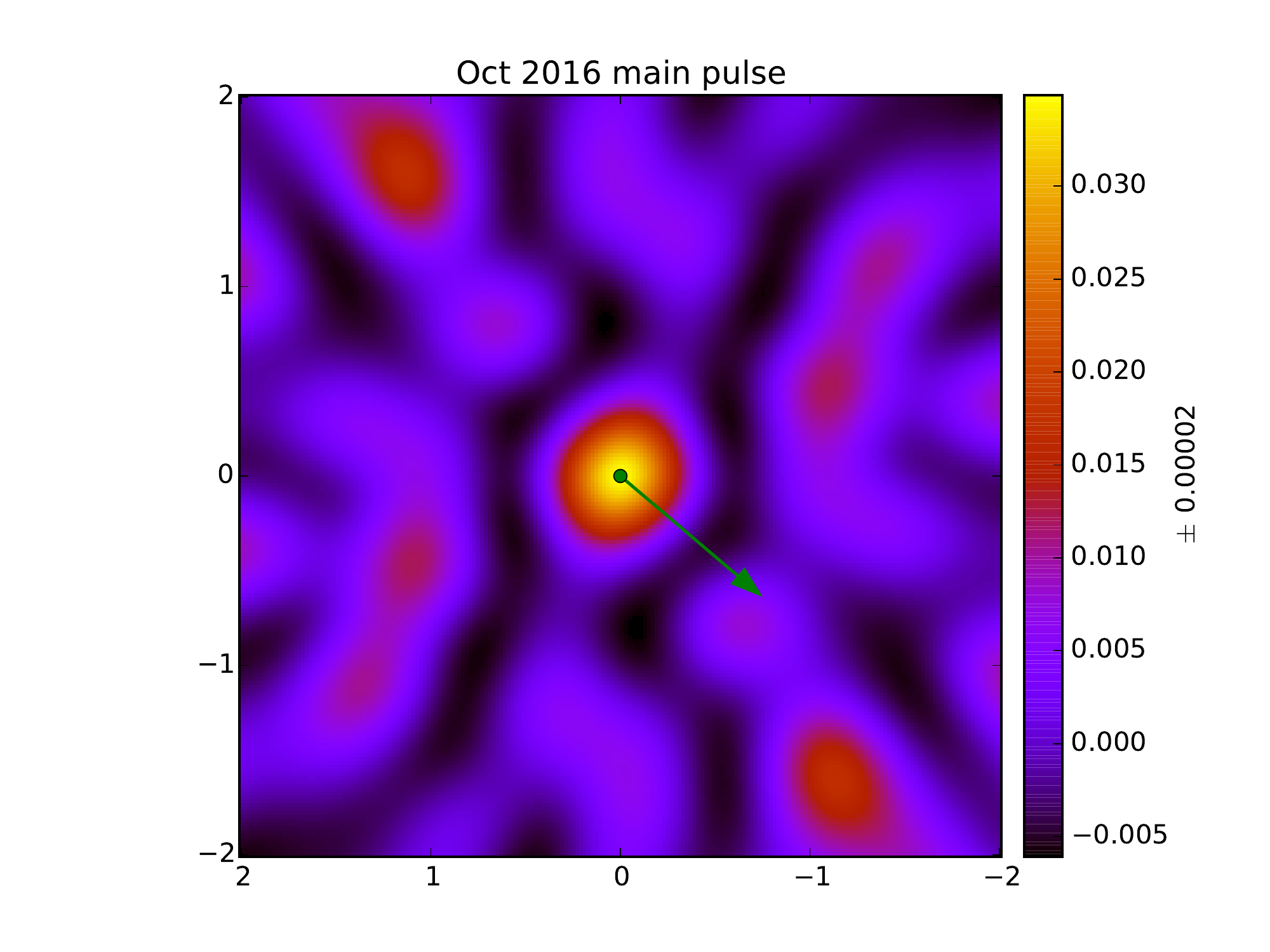}%
\includegraphics[width=0.29\textwidth,bb=80 18 560
  410,clip=true]{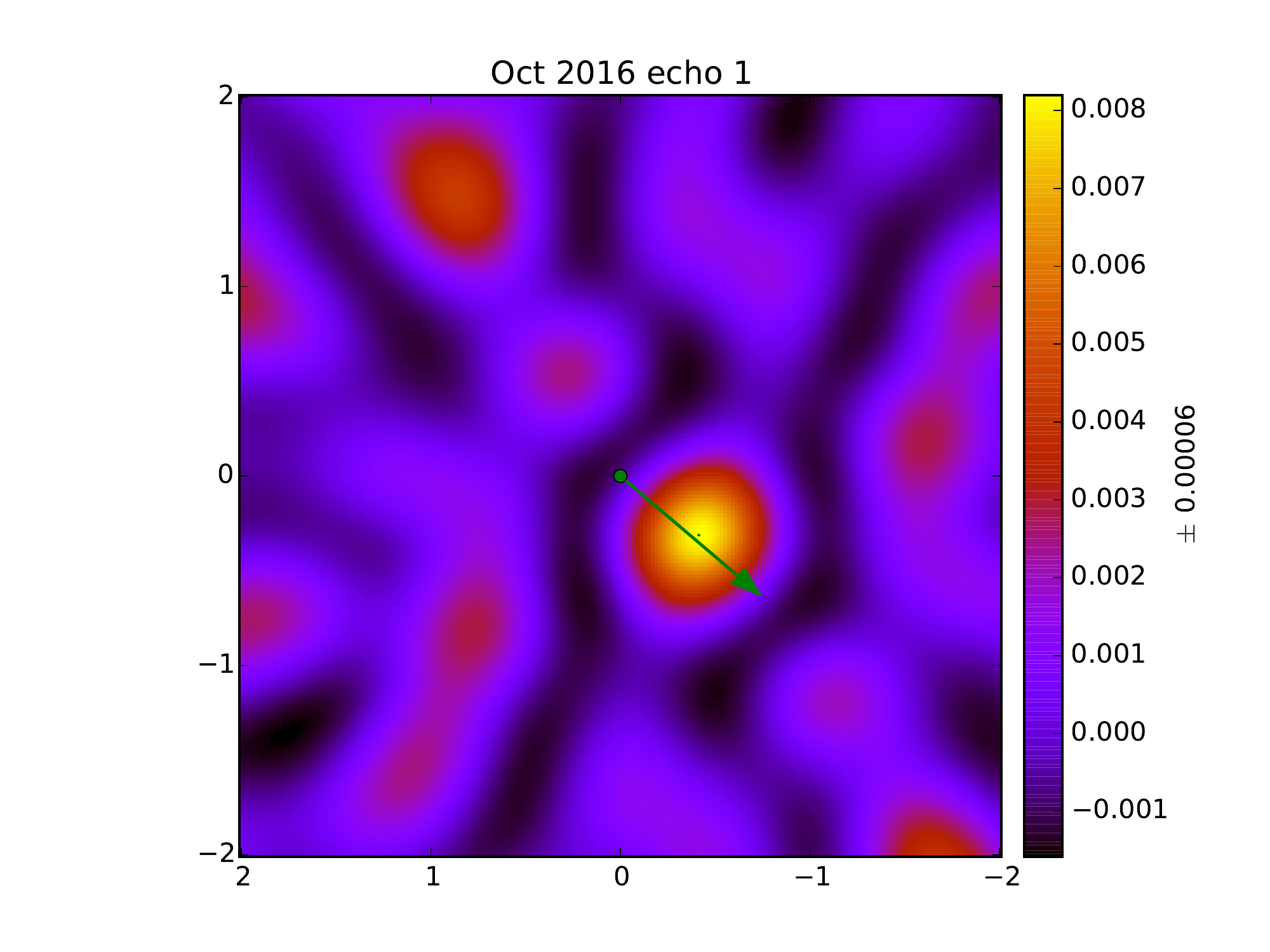}%
\includegraphics[width=0.29\textwidth,bb=80 18 560
  410,clip=true]{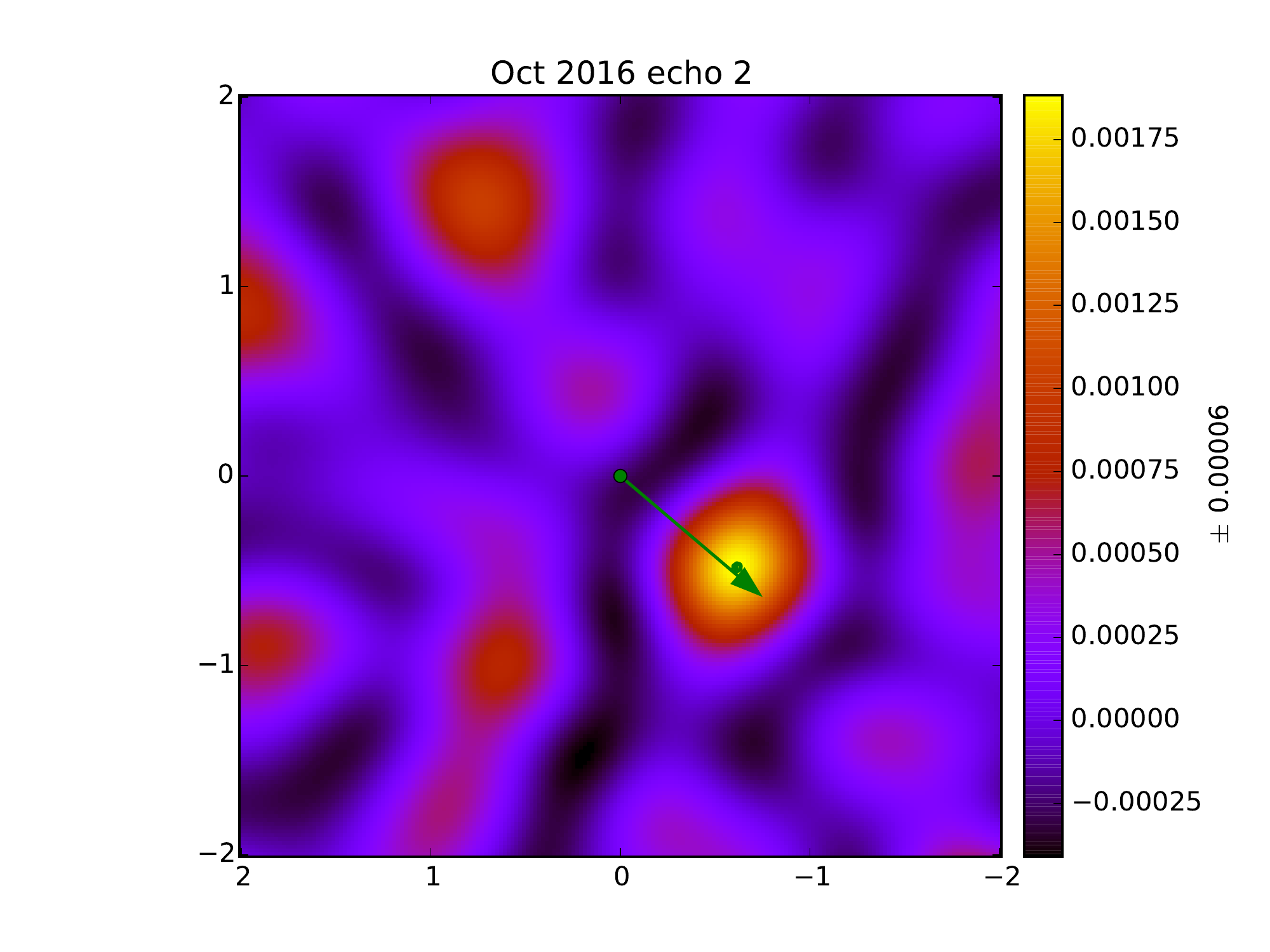} \\

\includegraphics[width=0.29\textwidth,bb=80 18 560
  410,clip=true]{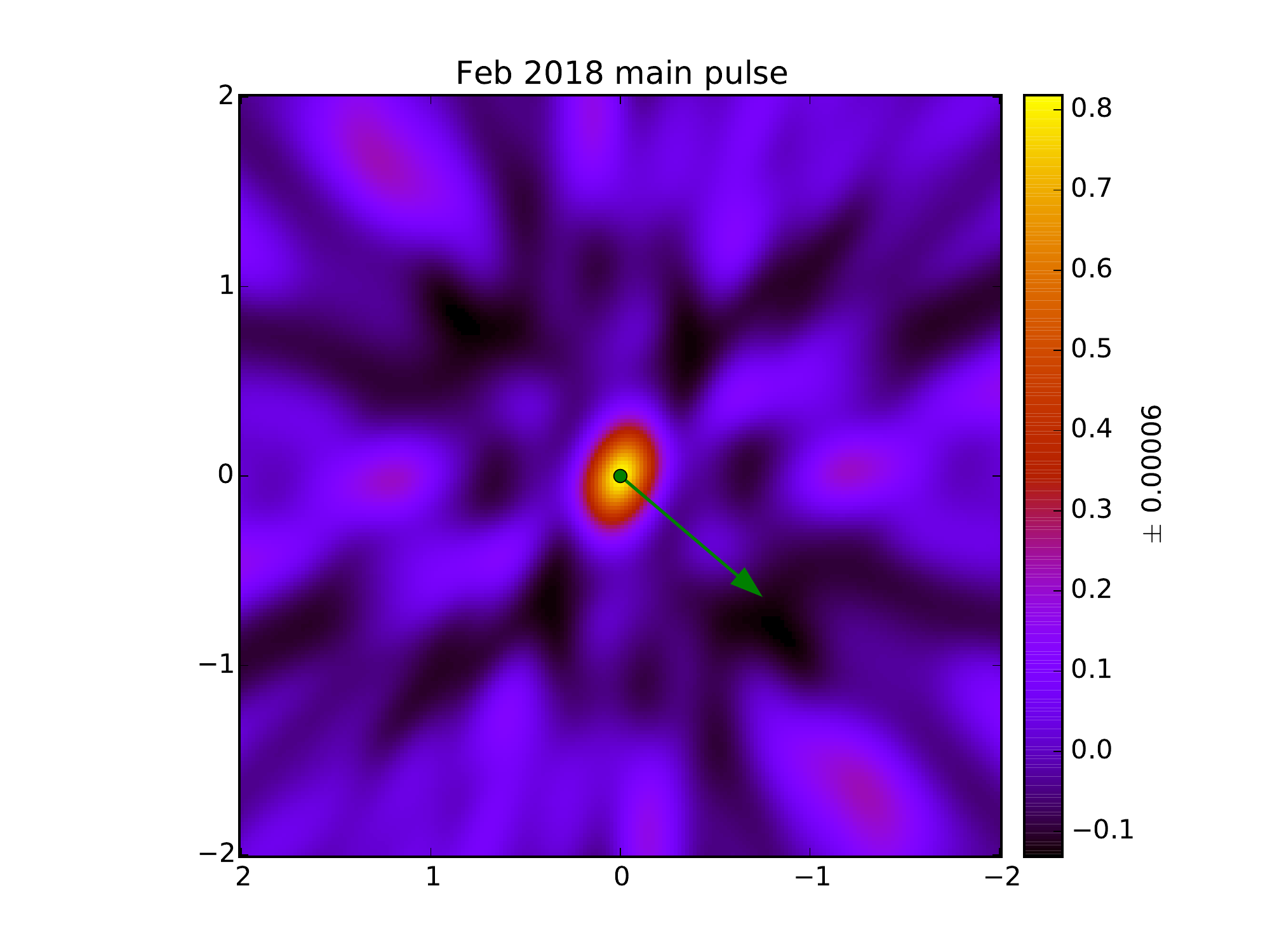}%
\includegraphics[width=0.29\textwidth,bb=80 18 560
  410,clip=true]{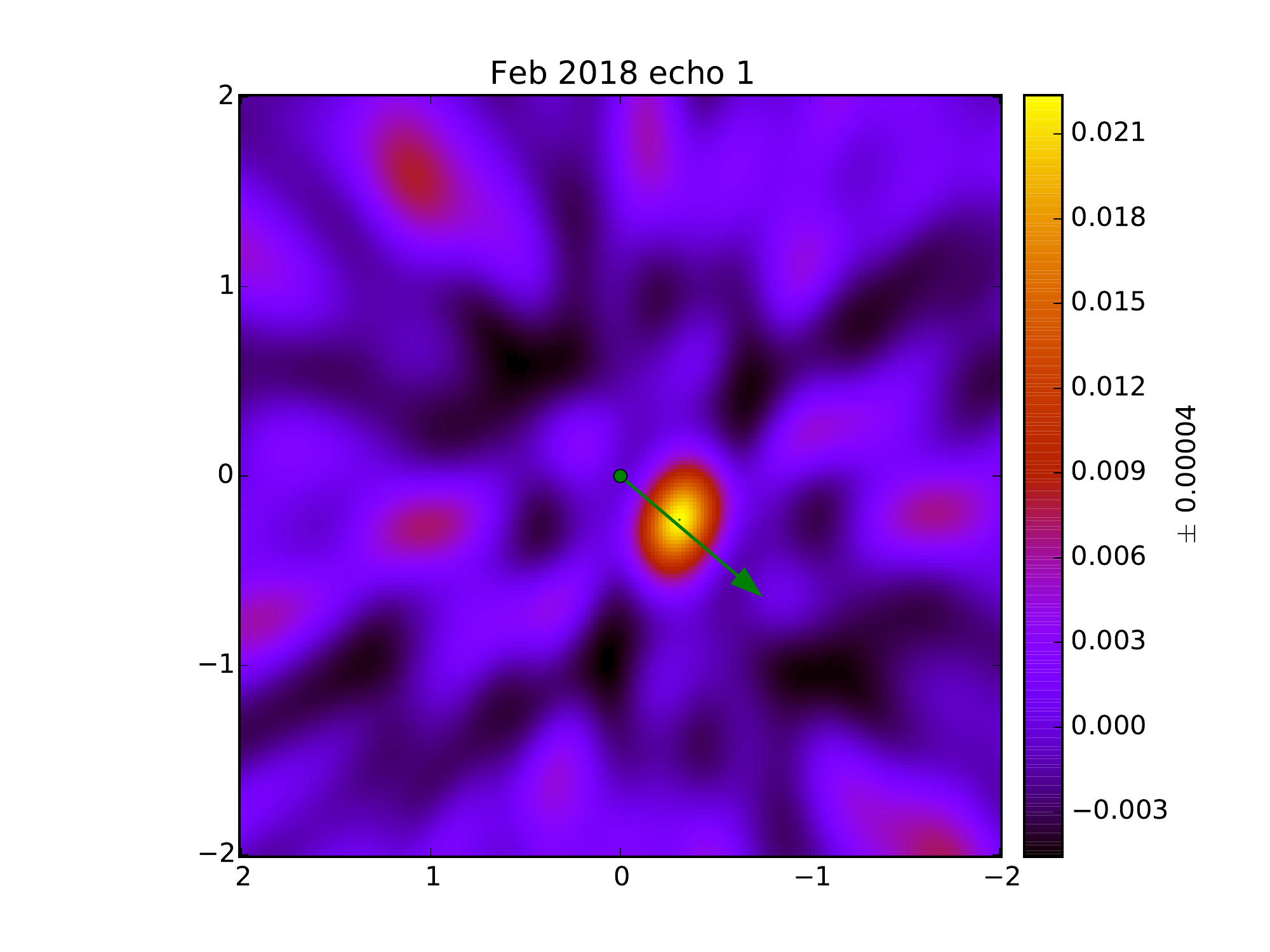}%
\includegraphics[width=0.29\textwidth,bb=80 18 560
  410,clip=true]{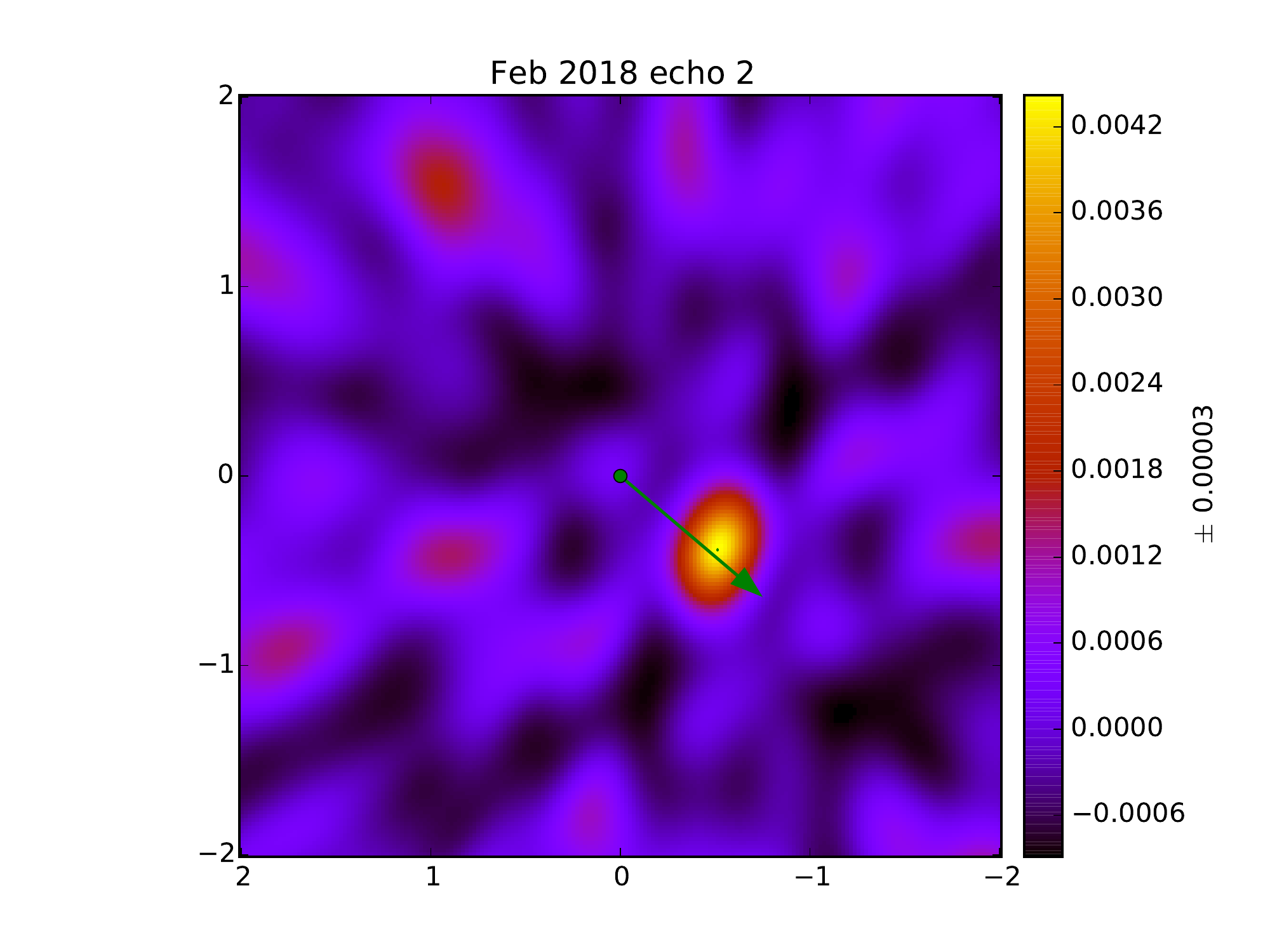}
\caption{Gated dirty images of the main pulse (left) and the first two
  echoes (centre and right). The top panels are from October 2016
  using GLOW only, the lower ones are with the international
  array. All features besides the main peak are sidelobes, the noise
  level is much lower. We notice the increased resolution with the
  full array. The arrow represents the pulsar proper motion of almost
  100\,mas/yr \citep{chatterjee09}. The alignment of the echo offsets
  with the proper motion seem to be accidental, even though some
  alignment is favoured by selection effects.}
  \label{fig:dirty}
\end{figure}

We continued observing the system more or less regularly, included
KAIRA and later also the other international LOFAR stations and the
core. Fig.~\ref{fig:dirty} shows dirty images from early GLOW and
later international data. We notice that the angular separation of the
echoes from the main pulse decreases with time in a way that is
consistent with the pulsar proper motion and fixed echo positions.

\section{Deconvolution}

The gated analysis presented so far is not optimal, because it is
based on the invalid assumption that the intrinsic profile does not
overlap with the echoes, and that delayed versions of the profile do
not overlap with each other. In order to solve this superposition
problem, we are developing a method to combine the angular
deconvolution (with the dirty beam) and the temporal deconvolution
(with the pulse profile) into one algorithm. The current version is a
generalisation of CLEAN. Standard CLEAN subtracts and collects
components by position (and implicitly flux). Our
delayed-profile-aware version of CLEAN uses the delay as additional
explicit parameter and replaces the flux by a spectrum as implicit
parameter. In this way the measured signal is decomposed into delayed
and shifted version of the intrinsic pulse profile, each potentially
with its own spectrum, because scattering is known the be strongly
chromatic.

\begin{figure}[hbt]
  \centering \includegraphics[width=0.8\textwidth]{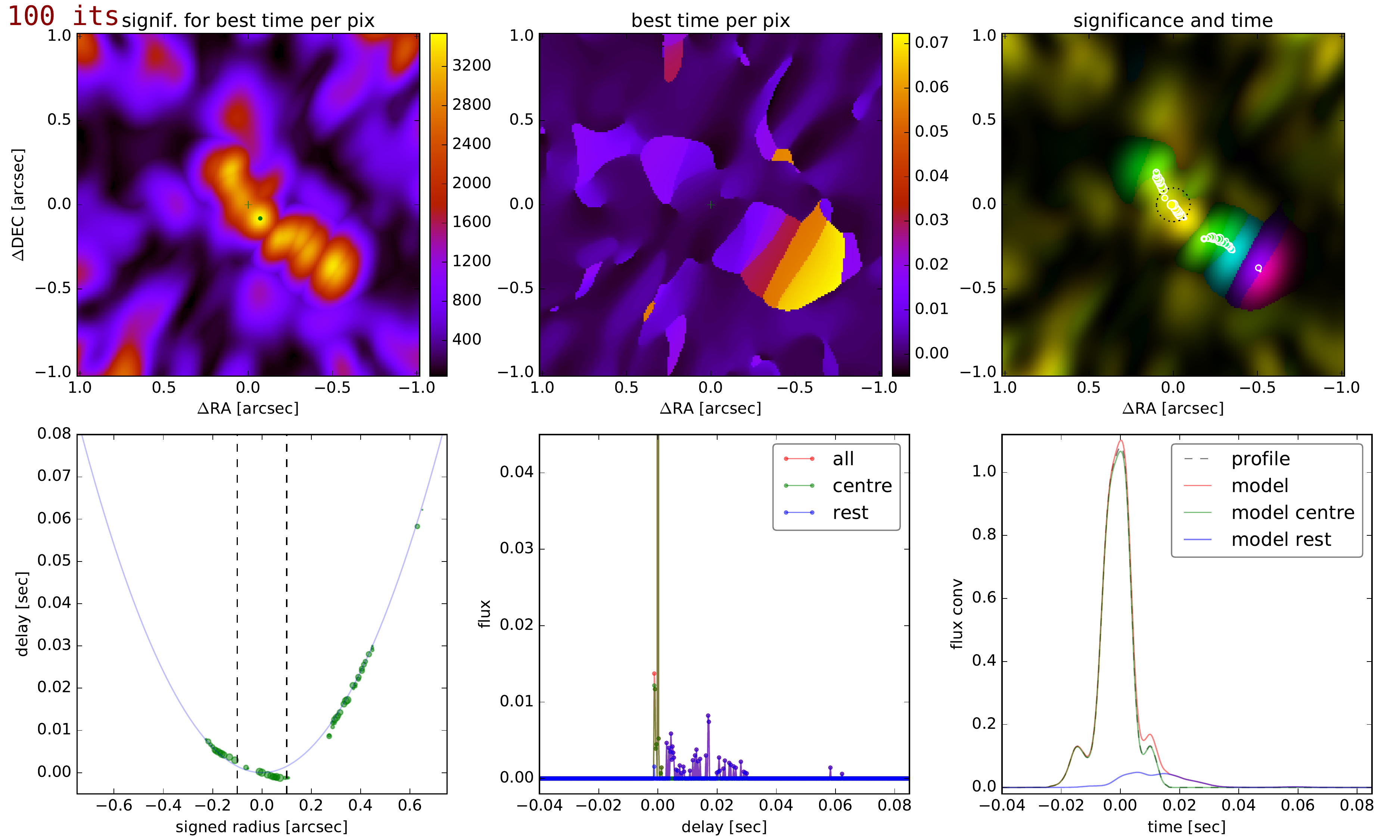}
  \caption{Intermediate results of our generalised CLEAN method after
    100 iterations.  Top: Residual map (significance as function of
    position, each for the optimum delay); optimum delay as function
    of position; the two combined (delay as colour, white circles show
    collected components). Bottom: Delay as function of signed radius
    (measured from upper left to bottom right) with expected parabola;
    components in delay space; the same convolved with the intrinsic
    profile and separated into central part and rest (echoes).}
  \label{fig:clean}
\end{figure}

Figure~\ref{fig:clean} illustrates the deconvolution process. The
method is not perfect, but it already separates the echoes from the
wings of the intrinsic profile very well (blue and green curves in
lower right panel). With simple gated imaging these components would
be merged. The echo components fit the expected parabola (lower left
panel for $D\sub d=117$\,pc) very well, even after more iterations.
The deviations in the central area are at least partly due to
inaccuracies of the deconvolution. This will be investigated further,
because real deviations from the parabola are direct measurements of
the delay happening in the scattering screen (in addition to the
geometric delay) and would be evidence for larger-scale lensing
effects.

\section{Discussion and outlook}

Low-frequency VLBI proves the nature of the ghosts in B1508+55 as
scattering-induced echoes and makes it possible to study the
scattering process in unprecedented detail.

The fact that the string of echoes directly crosses the pulsar is at
least consistent with a very anisotropic scattering that can deflect
only in this direction. There are components on both sides, which
means that some must have crossed the line of sight in the past, to
our knowledge without spectacular effects. This an important result in
itself, because large-scale lensing would behave differently. It
remains to be seen what crossings of the brighter components in the
future will show.

With the current monitoring programme we will be able to follow the
echo components in their evolution over time while the pulsar is
moving behind them. Their behaviour provides important information
about the density field that is causing the scattering.  The alignment
with the proper motion is still a puzzle, but it helps in accelerating
the evolution in this system.

Possible physical models are under investigation, e.g.\ variants of
ideas presented by \citet{walker17}. There is indeed an A2 star
1.37\,pc from the line of sight at a distance of $(120\pm8)\,$pc,
which is consistent with our estimate of the distance to the
scattering screen. Intriguingly the offset is also aligned almost
exactly with the line of echoes.

In the future we will also study other systems in the same way,
e.g.\ B2217+47, in which \citet{michilli18} also find strong evidence
for echoes.

\acknowledgments Many thanks go to everybody who helped with the
observations at all the stations, in particular (in alphabetic order)
Leszek B\l{}aszkiewicz, Tobia Carozzi, Julian Donner, Jean-Mathias
Grie\ss{}meier, Andreas Horneffer, Aris Karastergiou, Andrzej
Krankowski, J\"orn K\"unsem\"oller, Wojciech Lewandowski, Barbara
Matyjasiak, Derek McKay, Natasha Porayko, Mariusz Po\.z{}oga, Hanna
Rotkaehl, Tomasz Sidorowicz, Bartosz \'S{}mierciak, Marian Soida,
Caterina Tiburzi. Additional thanks go to Mark Walker and Artem
Tuntsov for discussions about their model.

\smallskip

{\footnotesize This work is based in part on results obtained with
  International LOFAR Telescope (ILT) equipment under project codes
  LC8\_008, LC9\_038, LC10\_002. LOFAR is the Low Frequency Array
  designed and constructed by ASTRON. It has observing, data
  processing, and data storage facilities in several countries, that
  are owned by various parties (each with their own funding sources),
  and that are collectively operated by the ILT foundation under a
  joint scientific policy. The ILT resources have benefitted from the
  following recent major funding sources: CNRS-INSU, Observatoire de
  Paris and Universit\'e d'Orl\'e{}ans, France; BMBF, MIWF-NRW, MPG,
  Germany; Science Foundation Ireland (SFI), Department of Business,
  Enterprise and Innovation (DBEI), Ireland; NWO, The Netherlands; The
  Science and Technology Facilities Council, UK; Ministry of Science
  and Higher Education, Poland.\\ In particular we made use of data
  from the Effelsberg (DE601) station funded by the
  Max-Planck-Gesellschaft; the Unterweilenbach (DE602) station funded
  by the Max-Planck-Institut f\"ur Astrophysik, Garching; the
  Tautenburg (DE603) station funded by the State of Thuringia,
  supported by the European Union (EFRE) and the Federal Ministry of
  Education and Research (BMBF) Verbundforschung project D-LOFAR I
  (grant 05A08ST1); the Potsdam (DE604) station funded by the
  Leibniz-Institut f\"ur Astrophysik, Potsdam; the J\"ulich (DE605)
  station supported by the BMBF Verbundforschung project D-LOFAR I
  (grant 05A08LJ1); and the Norderstedt (DE609) station funded by the
  BMBF Verbundforschung project D-LOFAR II (grant 05A11LJ1). The
  observations of the German LOFAR stations were carried out in the
  stand-alone GLOW mode (German LOng-Wavelength array), which is
  technically operated and supported by the Max-Planck-Institut f\"ur
  Radioastronomie, the Forschungszentrum J\"ulich and Bielefeld
  University. We acknowledge support and operation of the GLOW
  network, computing and storage facilities by the FZ-J\"ulich, the
  MPIfR and Bielefeld University and financial support from BMBF
  D-LOFAR III (grant 05A14PBA), and by the states of
  Nordrhein-Westfalia and Hamburg.  } \baselineskip1.8ex

{\small\bibsep0.ex \bibliographystyle{aa} \bibliography{proceedings}}

\end{document}